\documentclass[12pt]{article}

\usepackage{epsfig}
\usepackage{psfig}
\usepackage{latexsym}
\usepackage{axodraw}
\usepackage{amssymb}

\begin{document}

\begin{titlepage}

\baselineskip 24pt

\begin{center}

{\Large {\bf Lepton Transmutation in the Dualized Standard Model}}

\vspace{.5cm}

\baselineskip 14pt

{\large Jos\'e BORDES}\\
jose.m.bordes\,@\,uv.es\\
{\it Departament Fisica Teorica, Universitat de Valencia,\\
  calle Dr. Moliner 50, E-46100 Burjassot (Valencia), Spain}\\
\vspace{.2cm}
{\large CHAN Hong-Mo}\\
chanhm\,@\,v2.rl.ac.uk \\
{\it Rutherford Appleton Laboratory,\\
  Chilton, Didcot, Oxon, OX11 0QX, United Kingdom}\\
\vspace{.2cm}
{\large TSOU Sheung Tsun}\\
tsou\,@\,maths.ox.ac.uk\\
{\it Mathematical Institute, University of Oxford,\\
  24-29 St. Giles', Oxford, OX1 3LB, United Kingdom}

\end{center}

\vspace{.3cm}

\begin{abstract}

The successful explanation of fermion mixing and of the fermion mass
hierarchy by the Dualized Standard Model (DSM) scheme is based on the
premises of a fermion mass matrix rotating in generation space with 
changing scales at a certain speed, which could in principle lead to 
sizeable flavour-violation observable in high sensitivity experiments
such as BaBar.  However, a full perturbative calculation to 1-loop 
order reported here shows that this kinematical, flavour-violating
effect of a rotating mass matrix is off-set in the DSM by parallel effects
from rotating wave functions and vertices giving in the end only very 
small flavour-violations which are unlikely to be detectable by present 
experiments.  The result means that at least for the present the DSM 
scheme has survived yet another threat to its validity, which is indeed 
its most stringent and dangerous to-date.  It also provides some
clarification of certain concepts connected with the rotating mass 
matrix which had previously been found puzzling.

\end{abstract}

\end{titlepage}

\clearpage

\baselineskip 14pt

\setcounter{equation}{0}

\section{Introduction}

The Dualized Standard Model (DSM) \cite{dualgen} that we suggested has had 
some, to us, quite remarkable successes as a candidate solution to the 
fermion generation puzzle.  Apart from offering a {\it raison d'\^etre} 
for 3 generations of fermions, it gives a very simple explanation for such 
otherwise mysterious phenomena as fermion mixing and hierarchical fermion 
mass spectrum.  Indeed, with only 3 adjustable parameters the DSM scheme 
is able to reproduce already at the one-loop level the following quantities 
all within present experimental limits: the 3 mass ratios $m_c/m_t, m_s/m_b$
and $m_\mu/m_\tau$, all the 9 matrix elements $|V_{\alpha \beta}|$ of the 
CKM quark mixing matrix \cite{CKMmatrix}, plus the 2 elements $|U_{\mu 3}|$ 
and $|U_{e 3}|$ of the MNS lepton mixing matrix \cite{MNSmatrix} measured 
in neutrino oscillation experiments \cite{Superk,Soudan,Chooz}.  This means 
in particular that the DSM scheme automatically reproduces an empirical 
fact which has recently caught much attention and been much wondered at, 
namely that the angle $|U_{\mu 3}|$ as found in atmospheric neutrinos 
\cite{Superk,Soudan} is near maximal while the corresponding angles 
$|V_{ts}|$ and $|V_{cb}|$ for quarks are very small.  Furthermore, even 
for the other measured mass and mixing parameters, namely $m_u, m_d, m_e$, 
and $|U_{e 2}|$, which are beyond the reach of the one-loop calculation so 
far performed, the estimates obtained by extrapolation are nevertheless
quite sensible.  Altogether, the above quantities account for 12 of 
the Standard Model's twenty-odd paramters which have no explanation 
in the conventional formulation but are simply introduced as empirical 
inputs.  To us, this much agreement with experiment obtained with only 
3 adjustable parameters appear nontrivial \cite{phenodsm,revidsm}.

However, all these apparent ``successes'' of the DSM rely on the scheme's 
prediction that the fermion mass matrix should change its orientation 
in generation space (rotate) with changing energy scale in a prescribed 
manner.  That the fermion mass matrix should rotate with changing scale 
is not in itself special to the DSM, since even in the conventional 
formulation of the Standard Model the mass matrix will rotate by virtue 
of the renormalization group equation \cite{rge} so long as there is 
nontrivial mixing between the up and down fermion states \cite{Ramond}.  
But the speed of rotation so obtained is far below that required to 
derive the DSM results on fermion mass hierarchy and mixing summarized 
in the preceding paragraph \cite{impromat}.  Now, requiring a rotation 
of the mass matrix at a high speed could be dangerous for it could lead 
to sizeable flavour-violation in contradiction to experiment.  For instance, 
a rotating lepton mass matrix means that the lepton flavour states which 
are defined to be diagonal states of the mass matrix at some prescribed 
scale(s) will in general no longer remain diagonal states at a different 
energy scale, either of the mass matrix itself or of reaction amplitudes 
which depend on it.  Hence, as suggested in \cite{impromat}, in certain
reactions, leptons can change their flavours (transmute), simply by virtue 
of the kinematics of a rotating mass matrix even in the absence of a
flavour-changing vertex.  The amount of flavour-violation so induced 
depends on the speed at which the mass matrix rotates.  Specifically, 
according to \cite{transbhar}, a lepton mass matrix rotating at the 
speed demanded by DSM to give the results listed in the last paragraph 
could lead via kinematics alone to a cross section for the transmutation 
reaction: $e^+ e^- \rightarrow \mu^\pm \tau^\mp$ of as much as 80 fb 
at 10.58 GeV, where BaBar, for example, has collected already 20 
${\rm fb}^{-1}$ of data after only a year of running.  In other words, 
such a phenomenon, if real, could readily be observed in principle by 
an analysis of the existing BaBar data.  Hence, to test the validity of 
the DSM scheme, it is imperative to ascertain whether such effects are 
indeed obtained as predictions of the scheme.
    
The cited conclusion in \cite{transbhar}, however, does not mean that the 
DSM will necessarily lead to lepton flavour violations of such magnitude.
The calculation reported there evaluated only the kinematic effect of a 
rotating mass matrix considered in isolation without taking account of 
the dynamical mechanism driving it.  In the DSM scheme, however, the 
rotating mass matrix is deduced as a consequence of a specific driving
mechanism due to radiative corrections and this may give rise to other
rotation effects which may modify or even cancel the effect calculated 
in \cite{transbhar} from the rotating mass matrix.  Hence, to calculate 
properly in the DSM scheme the transmutation effects in say $e^+ e^-$ 
collisions, one should evaluate for consistency all radiative corrections 
to the amplitude to the same order.  In particular, to 1-loop order,
one should include one-loop insertions not just in the fermion propagator
to obtain the rotating mass matrix, but also in the external fermion lines 
(wave functions) and the interaction vertices.  

The purpose of this paper is, therefore, to perform a full perturbative 
calculation to one-loop order for lepton transmutation in $\gamma e$ 
and $e^+ e^-$ collisions in parallel to the calculations done in 
\cite{photrans,transbhar}.  We shall show that within the DSM framework 
as it is at present formulated, such a calculation can be unambiguously 
performed, all the relevant parameters in the scheme having already been 
determined by fitting the single-particle properties as detailed in
the first paragraph.  As a result of this calculation, we shall find 
that there are indeed scale-dependent rotation effects other than those 
deduced in \cite{photrans,transbhar} from the rotating mass matrix alone,
and that these rotation effects cancel exactly in 1-loop order, giving 
thus in total no scale-dependent transmutation.  In other words, despite 
the requirment in the DSM scheme of a sizeable rotation speed for the
mass matrix in order to explain fermion mass hierarchy and mixing, no
abnormally large flavour-violation is predicted.  The flavour-violating 
effects obtained from renormalization at 1-loop level are of the order 
$s/M^2$, where $s$ is the interaction energy and 
$M$ is the generic mass of the flavour bosons exchanged, 
and could thus be taken together with normal flavour-changing neutral
current (FCNC) effects which are suppressed if $M$ is large, as is
generally supposed to be the case.  Indeed, in an earlier investigation
on FCNC effects in the DSM framework \cite{fcnc}, $M$ was estimated to 
be of order 500 TeV which would give very small flavour-violations at 
energies available to present experiments.  It means therefore that
the effect as considered in \cite{impromat,photrans}, dangerous as it 
seemed at first sight, is unlikely to cause problem for the DSM.  

The conclusion of the present calculation also helps in elucidating some
basic but previously unclear concepts connected with fermion mixing and 
neutrino oscillations in the circumstances when the mass matrix rotates,
which elucidation has wider connotations beyond the context of the present 
paper.

\section{Preliminaries}

The DSM scheme makes use of a theoretical result previously derived that
there is in Yang--Mills theory a nonabelian version of electric--magnetic
duality \cite{dualsymm}, namely that dual to the gauge group $SU(N)$, 
there is a another local group $\widetilde{SU}(N)$ under which the theory 
is also symmetric, the potential of the latter group being related to 
the potential of the original gauge group via a nonabelian dual transform 
formulated in loop space which generalizes the Hodge star for the abelian 
theory.  When the theory is quantized, it was shown that the dual group
$\widetilde{SU}(N)$ is broken when the original gauge group $SU(N)$ is
confined \cite{dualcomm,tHooft}.  Hence, in the case of colour as in 
standard chromodynamics, there is, dual to colour $SU(3)$, automatically 
a broken 3-fold symmetry $\widetilde{SU}(3)$ which can play the role of
a ``horizontal group'' \cite{horizontal} for exactly 3 generations of 
fermions.  Furthermore, the framework offers natural candidates for the 
Higgs fields breaking the generation symmetry in the form of frame-vectors 
(complex dreibeins) in $\widetilde{SU}(3)$ space, suggesting thereby the 
manner in which the generation symmetry is broken.  The result is a highly 
predictive scheme for the description of fermion generations \cite{dualcons}.

Although in themselves conceptually interesting with perhaps much wider
implications, the assertions of the preceding paragraph concern us here
only in yielding a particular form of the Yukawa coupling and of the Higgs 
potential.  The suggested Yukawa coupling between Higgs and fermion fields 
takes the form:
\begin{equation}
\sum_{(a)[b]} Y_{[b]} \bar{\psi}_L^a  \phi^{(a)}_a \psi_R^{[b]} + h.c.,
\label{Yukawa}
\end{equation}
where $\psi_L^a, a = 1,2,3$ is the left-handed fermion field appearing as
a dual colour triplet, and $\psi_R^{([b])}$ are 3 right-handed fermion 
fields, each appearing as a dual colour singlet.  These are coupled to
3 triplets of Higgs fields $\phi^{(a)}_a, (a) = 1,2,3, a = 1,2,3$ each
being a space-time scalar.  Further, it was suggested that the Higgs 
fields themselves self-interact via the following potential:
\begin{equation}
V[\phi] = -\mu \sum_{(a)} |\phi^{(a)}|^2 + \lambda \left\{ \sum_{(a)}
   |\phi^{(a)}|^2 \right\}^2 + \kappa \sum_{(a) \neq (b)} |\bar{\phi}^{(a)}
   \cdot \phi^{(b)}|^2,
\label{Vofphi}
\end{equation}
for which a general vacuum can be expressed as:
\begin{equation}
\phi^{(1)} = \zeta \left( \begin{array}{c} x \\ 0 \\ 0 \end{array} \right);
\ \ \phi^{(2)} = \zeta \left( \begin{array}{c} 0 \\ y \\ 0 \end{array} \right);
\ \ \phi^{(3)} = \zeta \left( \begin{array}{c} 0 \\ 0 \\ z \end{array} \right),
\label{phivac}
\end{equation}
with 
\begin{equation}
\zeta = \sqrt{\mu/2\lambda},
\label{zeta}
\end{equation}
and $x, y, z$ all real and positive, satisfying:
\begin{equation}
x^2 + y^2 + z^2 = 1.
\label{xyznorm}
\end{equation}
Such a vacuum breaks the permutation symmetry of the $\phi$'s which is 
maintained in both (\ref{Yukawa}) and (\ref{Vofphi}), and also the 
$\widetilde{U(3)}$ 
gauge symmetry completely making thus all the vector gauge bosons in the 
theory massive by eating up all but 9 of the original 18 Higgs modes.

As a result, the tree-level mass matrix for each of the 4 fermion-species 
$T$ (i.e.\ whether of $U$- or $D$-type quarks, or of charged leptons $L$ or 
neutrinos $N$) is of the following form:
\begin{equation}
m = \tilde{m} \frac{1}{2} (1 + \gamma_5) + \tilde{m}^{\dagger} \frac{1}{2}
   (1 - \gamma_5),
\label{mtreep}
\end{equation}
where $\tilde{m}$ is a factorized matrix:
\begin{equation}
\tilde{m} \propto \left( \begin{array}{c} x \\ y \\ z \end{array} \right)
   (a, b, c),
\label{mtree}
\end{equation}
with $a, b, c$ being the Yukawa couplings $Y_{[b]}$.  By suitably redefining 
the right-handed fermion fields \cite{Weinberg} which in no way affects the 
physics, one can rewrite the mass matrix in the more convenient form with
no dependence on $\gamma_5$:
\begin{equation}
m = m_T \left( \begin{array}{c} x \\ y \\ z \end{array} \right) (x, y, z),
\label{mtreew}
\end{equation}
which is the form that we shall use in our calculations throughout.  We note
that this is of rank 1, having only one nonzero eigenvalue with eigenvector 
$(x, y, z)$ the components of which, being Higgs vev's, are independent of 
the fermion-species $T$.  Hence we have at the tree-level (i) that the 
fermion mass spectrum is `hierarchical' with one generation much heavier 
than the other two, (ii) that the CKM matrix giving the relative orientation 
between the eigenvectors of the up- and down-type fermions is the identity 
matrix.  

\begin{figure}[ht]
\begin{center}
{\unitlength=1.0 pt \SetScale{1.0} \SetWidth{1.0}
\begin{picture}(350,130)(0,0)

\Line(10,50)(120,50)
\DashCArc(65,50)(25,0,180){3}
\Text(0,50)[]{$p$}
\Text(130,50)[]{$p$}
\Text(65,85)[]{$k$}
\Text(65,0)[]{$(a)$}

\Photon(265,25)(265,75){3}{6.5}
\Line(230,100)(265,75)
\Line(300,100)(265,75)
\Text(220,100)[]{$p$}
\Text(310,100)[]{$p'$}
\DashCArc(265,75)(22,35.5,144.5){3}
\Text(265,115)[]{$k$}
\Text(265,0)[]{$(b)$}

\end{picture}}
\end{center}
\vspace*{5mm}
\caption{The relevant self-energy and vertex insertions}
\label{insert}
\end{figure}
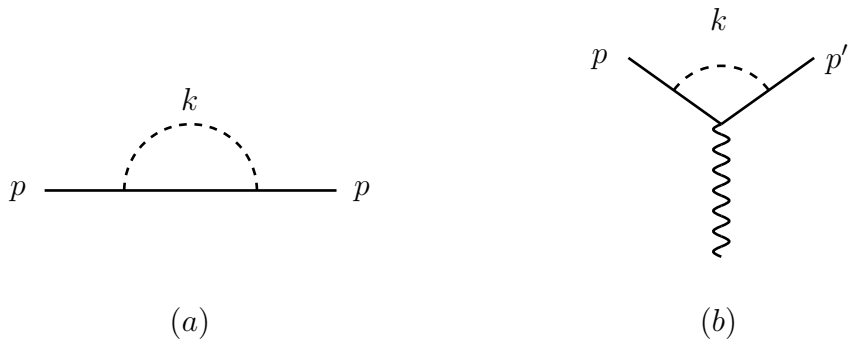

The results on fermion mass ratios and mixing parameters summarized at 
the beginning were obtained by the insertion of one (dual colour) Higgs 
loop to the fermion propagator as depicted in Figure \ref{insert}(a).
It has the effect of rotating the mass matrix (\ref{mtreew}) with changing
scale, hence giving nontrivial mixing between up and down states and 
nonvanishing masses to the lower generation fermion.  In order to 
calculate to the same loop order reaction processes by making all
possible insertions, we shall need the explicit mass spectrum of the 
Higgs fields and
their couplings to the fermions, which are deducible from (\ref{Vofphi})
and (\ref{Yukawa}) respectively.  

Following Weinberg \cite{Weinberg} we shall work in a real representation
where the remaining 9 Higgs bosons were found earlier \cite{ckm} to have
at tree-level the following mass values: 
\begin{eqnarray}
K = 1: & 8 \lambda \zeta^2 (x^2+y^2+z^2), \nonumber \\
K = 2: & 4 \kappa \zeta^2 (y^2 + z^2)   , \nonumber \\
K = 3: & 4 \kappa \zeta^2 (y^2 + z^2)   , \nonumber \\
K = 4: & 4 \kappa \zeta^2 (z^2 + x^2)   , \nonumber \\
K = 5: & 4 \kappa \zeta^2 (z^2 + x^2)   , \nonumber \\
K = 6: & 4 \kappa \zeta^2 (x^2 + y^2)   , \nonumber \\
K = 7: & 4 \kappa \zeta^2 (x^2 + y^2)   , \nonumber \\
K = 8: & 0                              , \nonumber \\
K = 9: & 0                              ,
\label{MK1to9}
\end{eqnarray}
and the following couplings to fermions:
\begin{equation}
\bar{\Gamma}_K = \bar{\gamma}_K \frac{1}{2} (1 + \gamma_5)
   + \bar{\gamma}^{\dagger}_K \frac{1}{2} (1 - \gamma_5),
\label{Gammabar}
\end{equation}
where
\begin{equation}
\bar{\gamma}_K = \rho |v_K \rangle \langle v_1|,
\label{gammabar}
\end{equation}
with $\rho$ the strength of the Yukawa coupling, and
\begin{eqnarray}
|v_1 \rangle & = & \left( \begin{array}{c} x \\ y \\ z \end{array} \right),
   \nonumber \\
|v_2 \rangle & = & \frac{1}{\sqrt{y^2 + z^2}}
   \left( \begin{array}{c} 0 \\ y \\ z \end{array} \right), \nonumber \\
|v_3 \rangle & = & \frac{i}{\sqrt{y^2 + z^2}}
   \left( \begin{array}{c} 0 \\ y \\-z \end{array} \right), \nonumber \\
|v_4 \rangle & = & \frac{1}{\sqrt{z^2 + x^2}}
   \left( \begin{array}{c} x \\ 0 \\ z \end{array} \right), \nonumber \\
|v_5 \rangle & = & \frac{i}{\sqrt{z^2 + x^2}}
   \left( \begin{array}{c} -x\\ 0 \\ z \end{array} \right), \nonumber \\
|v_6 \rangle & = & \frac{1}{\sqrt{x^2 + y^2}}
   \left( \begin{array}{c} x \\ y \\ 0 \end{array} \right), \nonumber \\
|v_7 \rangle & = & \frac{i}{\sqrt{x^2 + y^2}}
   \left( \begin{array}{c} x \\-y \\ 0 \end{array} \right),
\label{vK1to7}
\end{eqnarray}
while the two remaining (degenerate) ``zero modes'' can be assigned the 
following vectors orthogonal to $|v_1 \rangle$:
\begin{equation}
|v_8 \rangle = -\beta \left( \begin{array}{c} y-z \\ z-x \\ 
   x-y \end{array} \right);
|v_9 \rangle = \beta \left( \begin{array}{c} 1-x(x+y+z) \\ 1-y(x+y+z) \\ 
   1-z(x+y+z) \end{array} \right),
\label{vK8to9}
\end{equation}
with
\begin{equation}
\beta^{-2} = 3 - (x+y+z)^2.
\label{beta}
\end{equation}
These ``zero modes'' arise only by virtue of an ``accidental'' symmetry 
of the vacuum which is not present in the action itself and are thus 
unlikely to remain massless under radiative correction.

\begin{figure}[ht]
\begin{center}
{\unitlength=1.0 pt \SetScale{1.0} \SetWidth{1.0}
\begin{picture}(350,250)(0,0) 

\Line(35,50)(95,50)
\Photon(35,50)(5,80){-2}{5.5}
\Photon(95,50)(125,80){2}{5.5}
\Line(5,20)(35,50)
\Line(95,50)(125,20)
\DashCArc(35,50)(20,225,0){3}

\Line(235,50)(295,50)
\Photon(235,50)(205,80){-2}{5.5}
\Photon(295,50)(325,80){2}{5.5}
\Line(205,20)(235,50)
\Line(295,50)(325,20)
\DashCArc(295,50)(20,180,315){3}

\Line(135,180)(195,180)
\Photon(135,180)(105,210){-2}{5.5}
\Photon(195,180)(225,210){2}{5.5}
\Line(105,150)(135,180)
\Line(195,180)(225,150)
\DashCArc(165,180)(15,180,0){3}

\Line(-20,180)(40,180)
\Photon(-20,180)(-50,210){-2}{5.5}
\Photon(40,180)(70,210){2}{5.5}
\Line(-50,150)(-20,180)
\Line(40,180)(70,150)
\DashCArc(-35,165)(12,225,45){3}

\Line(290,180)(350,180)
\Photon(290,180)(260,210){-2}{5.5}
\Photon(350,180)(380,210){2}{5.5}
\Line(260,150)(290,180)
\Line(350,180)(380,150)
\DashCArc(365,165)(12,135,315){3}

\end{picture}}
\end{center}
\vspace*{5mm}
\caption{1-loop insertions in the amplitude for $\gamma e$ collision}
\label{egdiag}
\end{figure}
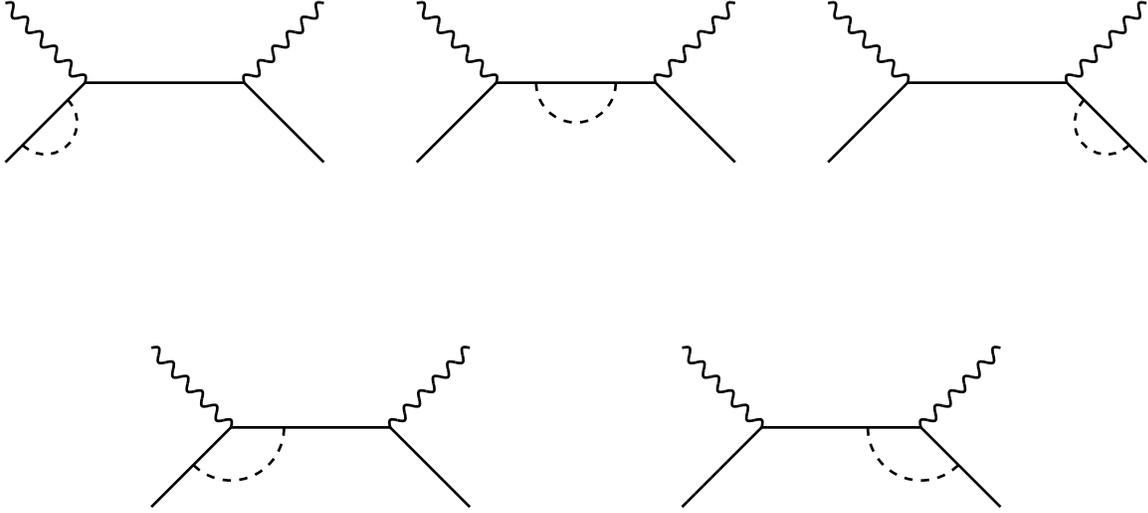

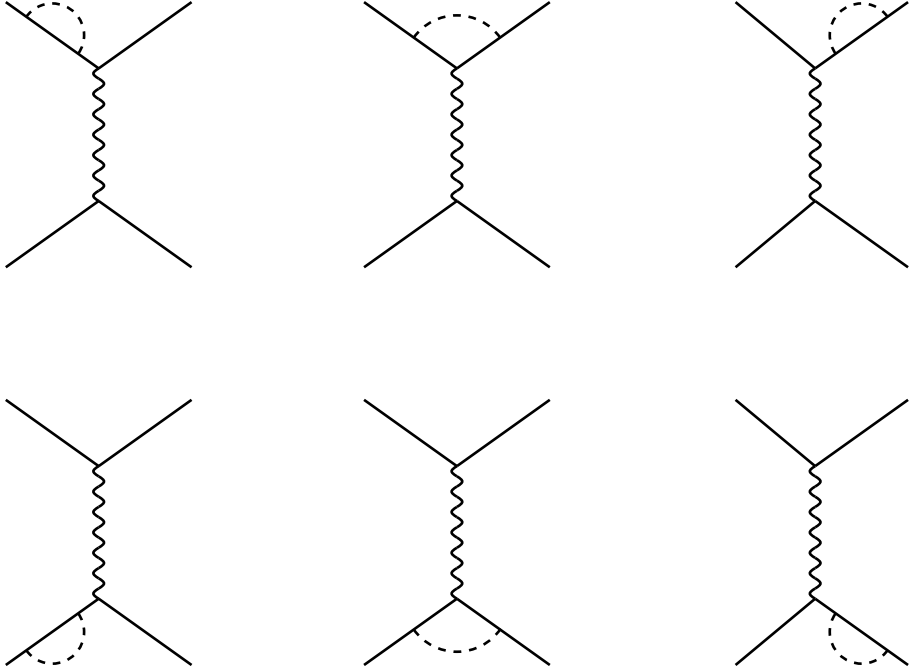
\begin{figure}[ht]
\begin{center}
{\unitlength=1.0 pt \SetScale{1.0} \SetWidth{1.0}
\begin{picture}(350,250)(0,0) 

\Photon(165,25)(165,75){2}{6.5}
\Line(165,75)(130,100)
\Line(130,0)(165,25)
\Line(165,25)(200,0)
\Line(200,100)(165,75)
\DashCArc(165,25)(20,215.5,325.5){3}

\Photon(30,25)(30,75){2}{6.5}
\Line(30,75)(-5,100)
\Line(-5,0)(30,25)
\Line(30,25)(65,0)
\Line(65,100)(30,75)
\DashCArc(12.5,12.5)(12,215.5,35.5){3}

\Photon(300,25)(300,75){2}{6.5}
\Line(300,75)(270,100)
\Line(270,0)(300,25)
\Line(300,25)(335,0)
\Line(335,100)(300,75)
\DashCArc(317.5,12.5)(12,145.5,325.5){3}

\Photon(165,175)(165,225){2}{6.5}
\Line(165,225)(130,250)
\Line(130,150)(165,175)
\Line(165,175)(200,150)
\Line(200,250)(165,225)
\DashCArc(165,225)(20,35.5,144.5){3}

\Photon(30,175)(30,225){2}{6.5}
\Line(30,225)(-5,250)
\Line(-5,150)(30,175)
\Line(30,175)(65,150)
\Line(65,250)(30,225)
\DashCArc(12.5,237.5)(12,325.5,145.5){3}

\Photon(300,175)(300,225){2}{6.5}
\Line(300,225)(270,250)
\Line(270,150)(300,175)
\Line(300,175)(335,150)
\Line(335,250)(300,225)
\DashCArc(317.5,237.5)(12,35.5,215.5){3}

\end{picture}}
\end{center}
\vspace*{5mm}
\caption{1-loop insertions in the amplitude for $e^+ e^-$ collision}
\label{eediag}
\end{figure}

With the above information, one can now proceed to evaluate transmutation
effects of scale-dependent rotation from 1-loop corrections to $e^+ e^-$ 
and $\gamma e$ reactions of present interest.  The main diagrams to be 
evaluated are listed respectively in Figures \ref{egdiag} and \ref{eediag}, 
plus the corresponding crossed diagrams, where we notice that since the
photon carries no generation (dual colour) index, it does not couple to
the (dual colour) Higgs fields, so that only 2 types of loop insertions 
occur, namely either self-energy insertions in the fermion line (internal
or external) or insertions in the fermion-fermion-photon vertex, as depicted
in Figure \ref{insert}.  Although the calculations for these insertions are
fairly standard, the fact that $m$ is a rotating matrix does cause some
complications which require consideration with due care.  We shall examine 
these 2 types of insertions in turn.\footnote{For notation we follow closely
\cite{Mandlshaw}; for matrix complications and other intricacies see
e.g.\  
\cite{Weinberg,ckm}.}

\section{The Self-Energy Insertion}

The fermion self-energy insertion of Figure \ref{insert}(a) takes the
form:
\begin{equation}
\Sigma(p) = \frac{i}{(4 \pi)^4} \sum_K \int d^4 k \frac{1}{k^2 - M_K^2}
   \bar{\Gamma}_K \frac{(p\llap/ - k\llap/) + m}{(p - k)^2 - m^2}
   \bar{\Gamma}_K,
\label{Sigma}
\end{equation}
with $m$ and $\bar{\Gamma}_K$ given in (\ref{mtreew}) and (\ref{Gammabar}).
Combining denominators by the standard Feynman parametrization and shifting 
the origin of the $k$-integration as usual, one obtains:
\begin{equation}
\Sigma(p) = \frac{i}{(4 \pi)^4} \sum_K \int_0^1 dx \bar{\Gamma}_K
   \left\{ \int d^4 k \frac{p\llap/ (1 -x) + m}{[k^2 - Q^2]^2} \right\}
   \bar{\Gamma}_K,
\label{Sigma1}
\end{equation}
with
\begin{equation}
Q^2 = m^2 x + M_K^2 (1 - x) - p^2 x(1 -x),
\label{Qsquare}
\end{equation}
where we note that $m$, being a matrix in generation space, cannot be
commuted through the coupling $\bar{\Gamma}_K$'s.  The integration over 
$k$ in (\ref{Sigma1}) is divergent and has to be regularized.   Following 
the standard dimensional regularization procedure, one obtains:
\begin{equation}
\Sigma(p) = - \frac{1}{16 \pi^2} \sum_K \int_0^1 dx \bar{\Gamma}_K
   \{\bar{C} - \ln (Q^2/\mu^2) \} [p\llap/ (1 - x) + m] \bar{\Gamma}_K,
\label{Sigma2}
\end{equation}
with $\bar{C}$ being the divergent constant:
\begin{equation}
\bar{C} = \lim_{d \rightarrow 4} \left[ \frac{1}{2 - d/2} - \gamma \right],
\label{Cbar}
\end{equation}
to be subtracted in the standard $\overline{\rm MS}$ scheme.

To extract the renormalized mass matrix:
\begin{equation}
m' = m + \delta m
\label{mprime}
\end{equation}
from $\Sigma(p)$, one normally puts in the denominator $p^2 = m^2$ and
commutes $p\llap/$ in the numerator to the left or right and replace by 
$m$ \cite{Weinberg}.  However, $m$ being now a matrix, this operation is 
a little more delicate.  In order to maintain the ``hermitan'', left-right
symmetric form (\ref{mtreep}) for the renormalized mass matrix $m'$, we
split the $p\llap/$ term into two halves, commuting half to the left and 
half to the right before replacing by $m$, and hence obtain for $\delta m$ 
the following:
\begin{eqnarray}
\delta m & = & \frac{\rho^2}{16 \pi^2} \sum_K \int_0^1 dx 
   \{ \bar{\gamma}_K m [\bar{C} - \ln (Q_0^2/\mu^2)] \bar{\gamma}_K
      \frac{1}{2}(1 + \gamma_5) \nonumber \\ 
   & & + \bar{\gamma}_K^{\dagger} m [\bar{C} - \ln (Q_0^2/\mu^2)]
      \bar{\gamma}_K^{\dagger} \frac{1}{2}(1 - \gamma_5) \} \nonumber \\
   & & + \frac{\rho^2}{32 \pi^2} \sum_K \int_0^1 dx (1 - x) 
      m \{ \bar{\gamma}_K^{\dagger} [\bar{C} - \ln (Q_0^2/\mu^2)]
      \bar{\gamma}_K \frac{1}{2}(1 + \gamma_5) \nonumber \\ 
   & & + \bar{\gamma}_K [\bar{C} - \ln (Q_0^2/\mu^2)]
      \bar{\gamma}_K^{\dagger} \frac{1}{2}(1 - \gamma_5) \} \nonumber \\
   & & + \frac{\rho^2}{32 \pi^2} \sum_K \int_0^1 dx (1 - x) 
      \{ \bar{\gamma}_K [\bar{C} - \ln (Q_0^2/\mu^2)]
      \bar{\gamma}_K^{\dagger} \frac{1}{2}(1 + \gamma_5) \nonumber \\ 
   & & + \bar{\gamma}_K^{\dagger} [\bar{C} - \ln (Q_0^2/\mu^2)]
      \bar{\gamma}_K \frac{1}{2}(1 - \gamma_5) \} m,
\label{deltam}
\end{eqnarray}
with
\begin{equation}
Q_0^2 = Q^2|_{p^2 = m^2} = m^2 x^2 + M_K^2 (1 - x).
\label{Qzerosquare}
\end{equation} 
This is what was evaluated\footnote{There is a sign error in 
the first term on the right of eq. (4.14) of \cite{ckm} due to a misprint in 
the formula for $\Sigma^{(\phi 1)}$ in eq. (3.2) quoted from \cite{Weinberg}
which means that the coefficient of the last term in eq. (5.8) of \cite{ckm}
should be $3/(64 \pi^2)$ instead of $5/(64 \pi^2)$.  Apart from the fact
that the 
numerical values given for the parameter
$\rho$ in eq. (6.8) should be increased by a 
factor $\sqrt{5/3}$, no other result given in that paper or in its sequels
such as \cite{phenodsm} is affected by this error.} in \cite{ckm} 
and is all that is 
needed to calculate the single-particle properties such as fermion mass 
and mixing parameters of interest to us there.  We notice in particular 
that the mass matrix $m'$ after renormalization rotates with changing 
scale $\mu$, which was the crucial property that gave rise in our earlier
papers \cite{ckm,phenodsm} to the distinctive fermion mass and mixing 
patterns observed in experiment.

For investigating transmutation processes, however, more information is 
needed, for which purpose we write $\Sigma(p)$ as:
\begin{equation}
\Sigma(p) = - \frac{\delta m}{\rho^2} + \frac{1}{2} (p\llap/ - m) B_L
   + \frac{1}{2} B_R (p\llap/ - m) + \Sigma_c(p),
\label{Sigma3}
\end{equation}
with
\begin{eqnarray}
B_L & = & - \frac{1}{16 \pi^2} \sum_K \int_0^1 dx (1 - x) 
   \{ \bar{\gamma}_K^{\dagger} [\bar{C} - \ln (Q_0^2/\mu^2)] 
   \bar{\gamma}_K \frac{1}{2}(1 + \gamma_5) \nonumber \\  
   & & + \bar{\gamma}_K [\bar{C} - \ln (Q_0^2/\mu^2)] \bar{\gamma}_K^{\dagger} 
   \frac{1}{2}(1 - \gamma_5) \}, \nonumber \\
B_R & = & - \frac{1}{16 \pi^2} \sum_K \int_0^1 dx (1 - x)
   \{ \bar{\gamma}_K [\bar{C} - \ln (Q_0^2/\mu^2)] 
   \bar{\gamma}_K ^{\dagger} \frac{1}{2}(1 + \gamma_5) \nonumber \\ 
   & & + \bar{\gamma}_K^{\dagger} [\bar{C} - \ln (Q_0^2/\mu^2)] \bar{\gamma}_K
   \frac{1}{2}(1 - \gamma_5) \},
\label{BLBR}
\end{eqnarray}
and
\begin{eqnarray}
\Sigma_c(p) & = & \frac{1}{16 \pi^2} \sum_K \int_0^1 \{ \bar{\gamma}_K m 
   \ln (Q^2/Q_0^2) \bar{\gamma}_K \frac{1}{2}(1 + \gamma_5) \nonumber \\
   & & + \bar{\gamma}_K^{\dagger} m \ln (Q^2/Q_0^2) \bar{\gamma}_K^{\dagger} 
   \frac{1}{2}(1 - \gamma_5) \} \nonumber \\
   & & +\frac{1}{16 \pi^2} \sum_K \int_0^1 (1 - x) 
   \{ \bar{\gamma}_K^{\dagger} p\llap/ \ln (Q^2/Q_0^2) 
   \bar{\gamma}_K \frac{1}{2}(1 + \gamma_5) \nonumber \\  
   & & + \bar{\gamma}_K p\llap/ \ln (Q^2/Q_0^2) \bar{\gamma}_K^{\dagger} 
   \frac{1}{2}(1 - \gamma_5) \}.
\label{Sigmac} 
\end{eqnarray}
We note that $\Sigma_c$ so extracted is both finite and independent of 
the renormalization scale $\mu$.  Indeed for those terms in the sum over
$K$ for which $M_K$ is large, the contribution is only of order $s/M_K^2$,
as can be seen by writing:
\begin{equation}
\ln (Q^2/Q_0^2) = \ln \left[ 1 + \frac{(m^2 - p^2) x (1 - x)}{m^2 x^2 
   + M_K^2 (1 - x)} \right]
\label{lnQQ0}
\end{equation}
which for $0 \leq x \leq 1$ is, for large $M_K$, $\leq (m^2 - p^2)/M_K^2
\sim s/M_K^2$.

The insertion (\ref{Sigma3}) when added to an internal fermion line thus
gives:
\begin{equation}
\frac{1}{p\llap/ - m} \longrightarrow \frac{1}{p\llap/ - m'}
   - \frac{\rho^2}{2} B_L \frac{1}{p\llap/ - m}
   - \frac{\rho^2}{2} \frac{1}{p\llap/ - m} B_R
   - \rho^2 \frac{1}{p\llap/ - m} \Sigma_c(p) \frac{1}{p\llap/ - m},
\label{internalp}
\end{equation}
and when added to an external fermion line:
\begin{eqnarray}
u(p) & \longrightarrow & u'(p) - \frac{\rho^2}{2} B_L u(p) 
   - \rho^2 \frac{1}{p\llap/ - m} \Sigma_c(p) u(p), \\
\bar{u}(p) & \longrightarrow & \bar{u}'(p) - \frac{\rho^2}{2} \bar{u}(p) B_R
   - \rho^2 \bar{u}(p) \Sigma_c(p) \frac{1}{p\llap/ - m},
\label{externalp}
\end{eqnarray}
where $u'(p)$ is a solution of the Dirac equation with the renormalized
mass matrix $m'$:
\begin{equation}
(p\llap/ - m') u'(p) = 0.
\label{Diracp}
\end{equation}
These conclusions follow closely those in e.g.\ ordinary QED apart from that,
$m$ being a matrix and therefore noncommuting, (i) $B_L$ and $B_R$ are
different so that $u(p)$ and $\bar{u}(p)$ are renormalized differently,
(ii) the finite part $\Sigma_c(p)$ applying on $u(p)$ or $\bar{u}(p)$ does
not necesssarily give zero.  One notes also that $u(p)$ or $\bar{u}(p)$ 
picks up automatically just one-half of the $B$ contribution, i.e.\ either
$B_L/2$ or $B_R/2$, without the usual argument with adiabatic switching on
and off of the interaction being invoked.

\section{The Vertex Insertion}

Next, the vertex insertion of Figure \ref{insert}(b) takes the form:
\begin{equation}
\Lambda^\mu(p,p') = - \frac{i}{(2 \pi)^4} \sum_K \int d^4 k 
   \frac{1}{k^2 - M_K^2} \bar{\Gamma}_K \frac{(p\llap/' - k\llap/) + m}
   {(p' - k)^2 - m^2} \gamma^\mu \frac{(p\llap/ - k\llap/) + m}
   {(p - k)^2 - m} \bar{\Gamma}_K.
\label{Lambdamu}
\end{equation}
Combining denominators as usual with the Feynman parametrization, then
shifting the origin of the $k$-integration and dropping terms odd in $k$,
one obtains:
\begin{equation}
\Lambda^\mu(p,p') = - \frac{2 i}{(2 \pi)^4} \sum_K \int_0^1 dx \int_0^x dy
   \int d^4 k\,\bar{\Gamma}_K \frac{k\llap/ \gamma^\mu k\llap/}{[k^2 - P^2]^3}
   \bar{\Gamma}_K + \Lambda_c^\mu(p,p'),
\label{Lambdamu2}
\end{equation}
with
\begin{equation}
\Lambda_c^\mu(p,p') 
   = -\frac{2 i}{(2 \pi)^4} \sum_K \int_0^1 dx \int_0^x dy \int d^4 k
   \,\bar{\Gamma}_K \frac{{\cal N}}{[k^2 - P^2]^3} \bar{\Gamma}_K,
\label{Lambdacmu}
\end{equation}
\begin{equation}
{\cal N} = [p\llap/'(1 - x + y) - p\llap/(1 -x) +m] \gamma^\mu
   [p\llap/x - p\llap/'(x - y) +m],
\label{calN}
\end{equation}
and
\begin{equation}
P^2 = m^2(1 - y) + M_K^2 y - p^2 x(1 - x) - p'^2 (x - y)(1 - x + y)
   + 2pp'(1 - x)(x - y),
\label{Psquare}
\end{equation}
where $\Lambda_c^\mu(p,p')$, we note, is convergent, scale-independent, and
of order $s/M_K^2$ for large $M_K$.

The divergent integral over $k$ in (\ref{Lambdamu2}) we regularize again
by dimensional regularization obtaining an answer which we choose to write
as:
\begin{equation}
\Lambda^\mu(p,p') = \frac{1}{2} \gamma^\mu L_L + \frac{1}{2} L_R \gamma^\mu
   + \Lambda_c^\mu(p,p'),
\label{Lambdamu3}
\end{equation}
with
\begin{eqnarray}
L_L & = & - \frac{1}{16 \pi^2} \sum_K \int_0^1 dx \int_0^x dy\; \{ 
   \bar{\gamma}_K^{\dagger} [\bar{C} - \ln (P^2/\mu^2)] \bar{\gamma}_K
   \frac{1}{2}(1 + \gamma_5) \nonumber \\
   & & + \bar{\gamma}_K [\bar{C} - \ln (P^2/\mu^2)] \bar{\gamma}_K^{\dagger}
   \frac{1}{2}(1 - \gamma_5) \}, \nonumber \\
L_R & = & - \frac{1}{16 \pi^2} \sum_K \int_0^1 dx \int_0^x  dy \;\{ 
   \bar{\gamma}_K [\bar{C} - \ln (P^2/\mu^2)] \bar{\gamma}_K^{\dagger}
   \frac{1}{2}(1 + \gamma_5) \nonumber \\
   & & + \bar{\gamma}_K^{\dagger} [\bar{C} - \ln (P^2/\mu^2)] \bar{\gamma}_K
   \frac{1}{2}(1 - \gamma_5) \},
\label{LLLR}
\end{eqnarray}
where $\bar{C}$ is again the divergent constant in (\ref{Cbar}).

We notice that $L_L$ and $L_R$ in (\ref{LLLR}) are very similar to $B_L$ 
and $B_R$ in (\ref{BLBR}) obtained in the self-energy insertion.  Indeed,
if we take the difference $B_L - L_L$ ($B_R - L_R$), the divergent and
scale dependent parts cancel leaving only a term proportional to:
\begin{equation}
\Delta I = \int_0^1 dx (1 - x) \ln (Q_0^2) - \int_0^1 dx \int_0^x dy \ln (P^2),
\label{DeltaI}
\end{equation}
which we shall show is of order $s/M_K^2$ for $M_K$ large.  This is not
surprising since the self-energy and vertex insertions are related by 
the Ward identity:
\begin{equation}
\frac{\partial \Sigma(p)}{\partial p^\mu} = \Lambda^\mu(p,p),
\label{Wardid}
\end{equation}
which when applied to the formulae (\ref{Sigma3}) and (\ref{Lambdamu3})
would suggest such a result.  To see this explicitly, we note first that
for $s/M_K^2 \ll 1$, and $ 0 \leq x \leq 1, 0 \leq y \leq x$, one can
approximate $P^2$ as:
\begin{equation}
P^2 \sim M_K^2 y - (p - p')^2 x (1 - x) + m^2.
\label{Papprox}
\end{equation}
Secondly, we recall that the second integral on the right of (\ref{DeltaI})
is only symbolic, because the expression
\begin{equation}
\int_0^x dy\,[\bar{C} - \ln (P^2)]
\end{equation}
really means
\begin{equation}
\lim_{d \rightarrow 4} \int_0^x dy\,\Gamma(2 - d/2)\,(P^2)^{d/2 - 2},
\label{properlim}
\end{equation}
where  the
integral over $y$ should be performed first before taking the limit
for the proper regularization procedure, giving the estimate:
\begin{equation}
   \bar{C} x - x \ln (M_K^2 x + m^2),
\label{properlim2}
\end{equation}
which when substituted in place of the integral over $y$ in (\ref{DeltaI})
is easily seen to give for (\ref{DeltaI}) a value of order $s/M_K^2$ for
large $M_K$ as claimed.

With these observations for the vertex insertion in addition to those
above for the self-energy insertion, we are now in a position to consider
lepton-transmutations in $e^+ e^-$ and $\gamma e$ collisions.

\section{Transmutation in $\gamma e$ and $e^+ e^-$}

By transmutation here, we mean a reaction which violates 
flavour-conserva\-tion
by virtue of rotation effects (in generation space) under changes of 
scale which render the reaction amplitude non-diagonal in the flavour
states.  In the DSM scheme, this comes about mainly through loop diagrams
with (dual colour) Higgs exchange.  Indeed, it was the insertion of 
Figure \ref{insert}(a) into the fermion propagator which gave rise to the
rotating mass matrix in the first place and led to the DSM explanation
of fermion mixing and the fermion mass hierarchy \cite{ckm}.  There were
other insertions which gave mass matrix rotations, such as (dual colour)
gauge boson loops and tadpoles, but these were shown to give only effects of 
much lower magnitude so as to be negligible for present purposes.  In this 
paper, therefore, we shall be restricted to only 1-(dual colour)-Higgs-loop
diagrams.
 
\begin{figure}[ht]
\begin{center}
{\unitlength=1.0 pt \SetScale{1.0} \SetWidth{1.0}
\begin{picture}(350,100)(0,0) 
\Line(30,50)(100,50)
\Photon(30,50)(0,80){-2}{5.5}
\Photon(100,50)(130,80){2}{5.5}
\Line(0,20)(30,50)
\Line(100,50)(130,20)

\Text(-10,20)[]{$l_\alpha$}
\Text(140,20)[]{$l_\beta$}
\Text(-10,80)[]{$\gamma$}
\Text(140,80)[]{$\gamma$}
\Text(65,0)[]{$(a)$}

\Line(230,50)(300,50)
\Photon(230,50)(320,80){2}{9.5}
\Photon(300,50)(210,80){-2}{9.5}
\Line(200,20)(230,50)
\Line(300,50)(330,20)
\Text(190,20)[]{$l_\alpha$}
\Text(340,20)[]{$l_\beta$}
\Text(190,80)[]{$\gamma$}
\Text(340,80)[]{$\gamma$}
\Text(265,0)[]{$(b)$}
\end{picture} }
\end{center}
\caption{Tree diagrams for the reaction $\gamma \ell_\alpha \longrightarrow 
\gamma \ell_\beta$}
\label{egdiagtr}
\end{figure}
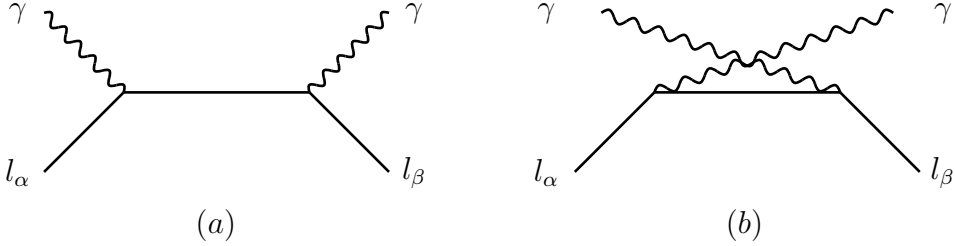

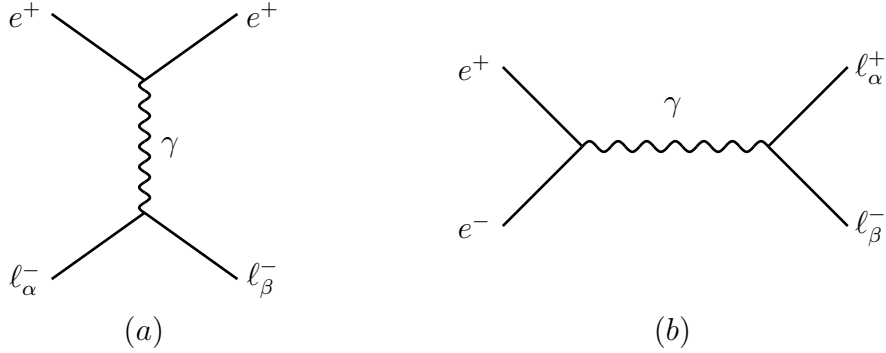
\begin{figure}[ht]
\begin{center}
{\unitlength=1.0 pt \SetScale{1.0} \SetWidth{1.0}

\begin{picture}(350,130)(0,0)

\Photon(65,25)(65,75){2}{6.5}
\Line(65,75)(30,100)
\Line(30,0)(65,25)
\Line(65,25)(100,0)
\Line(100,100)(65,75)

\Text(20,0)[]{$\ell_\alpha^-$}
\Text(20,100)[]{$e^+$}
\Text(110,100)[]{$e^+$}
\Text(110,0)[]{$\ell_\beta^-$}
\Text(75,50)[]{$\gamma$}
\Text(65,-20)[]{$(a)$}

\Photon(230,50)(300,50){2}{6.5}
\Line(230,50)(200,80)
\Line(330,80)(300,50)
\Line(200,20)(230,50)
\Line(300,50)(330,20)

\Text(190,20)[]{$e^-$}
\Text(340,20)[]{$\ell_\beta^-$}
\Text(190,80)[]{$e^+$}
\Text(340,80)[]{$\ell_\alpha^+$}
\Text(265,65)[]{$\gamma$}
\Text(265,-20)[]{$(b)$}
\end{picture}}
\end{center}
\vspace*{5mm}
\caption{Tree diagrams for the reaction $e^+ e^- \longrightarrow 
\ell_\alpha^+ \ell_\beta^-$}
\label{eediagtr}
\end{figure}

For the two reactions under consideration, one starts then with a mass 
matrix $m$ diagonal in the lepton-flavour states $\tau, \mu$ and $e$ 
giving for the reaction:
\begin{equation}
\gamma \ell_\alpha \longrightarrow \gamma \ell_\beta,
\label{egamma}
\end{equation}
at tree level the diagrams in Figure \ref{egdiagtr}, and for the reaction:
\begin{equation}
e^+ e^- \longrightarrow \ell_\alpha^+ \ell_\beta^-,
\label{eebar}
\end{equation}
the tree-level diagrams in Figure \ref{eediagtr}.  Explicitly, the
reaction amplitudes are, ignoring numerical factors:
\begin{equation}
\bar{u}(p') \gamma^\mu \frac{i}{(p\llap/ + k\llap/) - m} \gamma_\mu u(p),
\label{ampegtr}
\end{equation}
for the diagram of Figure \ref{egdiagtr}(a), and:
\begin{equation}
[\bar{u}(p') \gamma^\mu u(p)] \frac{1}{(p'-p)^2} [\bar{v}(q) \gamma_\mu
   v(q')],
\label{ampeetr}
\end{equation}
for the diagram of Figure \ref{eediagtr}(a), with similar formulae for
the diagrams (b) in each case.  The mass matrix $m$ being diagonal in 
the flavour states, so also are the reaction amplitudes, giving thus no 
flavour-violation at this tree-level.

\begin{figure}[ht]
\vspace*{5mm}
\begin{center}
{\unitlength=1.0 pt \SetScale{1.0} \SetWidth{1.0}

\begin{picture}(350,100)(0,0) 
\Line(30,50)(100,50)
\Photon(30,50)(0,80){-2}{5.5}
\Photon(100,50)(130,80){2}{5.5}
\Line(0,20)(30,50)
\Line(100,50)(130,20)
\DashLine(15,35)(115,35){4}

\Text(65,-20)[]{$(a)$}

\Photon(265,25)(265,75){2}{6.5}
\Line(265,75)(230,100)
\Line(230,0)(265,25)
\Line(265,25)(300,0)
\Line(300,100)(265,75)
\DashLine(282.5,87.5)(282.5,12.5){4}

\Text(265,-20)[]{$(b)$}

\end{picture} }
\end{center}
\caption{Convergent 1-loop diagrams}
\label{benchfig}
\end{figure}
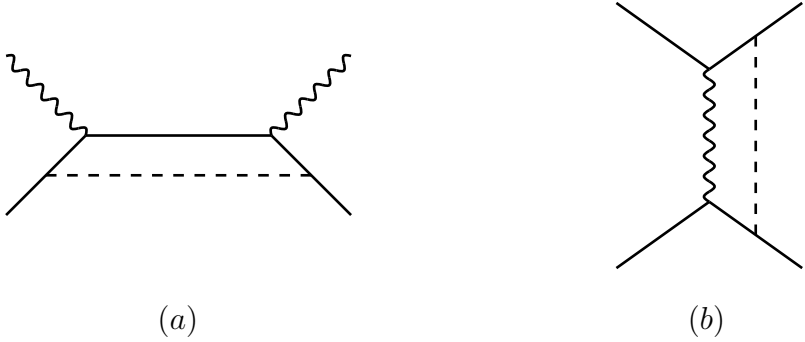

To calculate now the amplitude for the reaction (\ref{egamma}) to 1-loop
order, we have to add to Figure \ref{egdiagtr} all 1-loop diagrams with
(dual colour) Higgs exchange, i.e.\ the diagrams in Figure \ref{egdiag} 
with the fermion self-energy and vertex insertions studied in the preceding 
two sections, plus the diagram of Figure \ref{benchfig}(a).  This last-named
diagram is finite and is easily seen to give only effects of order $s/M_K^2$.
Adding the other diagrams and making use of the results in (\ref{internalp}),
(\ref{externalp}), and (\ref{Lambdamu3}), one obtains to order $\rho^2$ 
the result:
\begin{eqnarray}
& & \bar{u}'(p') \gamma^\mu \frac{i}{(p\llap/ + k\llap/) - m'} 
   \gamma_\mu u'(p) \nonumber\\
   & + & \frac{\rho^2}{2} \bar{u}(p') [\gamma^\mu (L_L - B_L) + (L_R - B_R) 
   \gamma^\mu] \frac{i}{(p\llap/ + k\llap/) - m} \gamma_\mu u(p) \nonumber\\
   & + & \frac{\rho^2}{2} \bar{u}(p') \gamma^\mu \frac{i}{(p\llap/ + k\llap/) 
   - m} [\gamma_\mu (L_L - B_L) + (L_R - B_R) \gamma_\mu] u(p)
\label{ampegren}
\end{eqnarray}
plus terms involving the quantities $\Sigma_c$ and $\Lambda_c^\mu$ which 
we recall are of order $s/M_K^2$ for large $M_K$.  We recall further that 
in the differences $B_L - L_L$ and $B_R - L_R$, the divergent and
the scale dependent parts both cancel, leaving in each only a finite 
part which
is again of order $s/M_K^2$ for large $M_K$.  Hence, if $M_K$ is indeed
large for all $K$, then the renormalized amplitude (\ref{ampegren}) will
reduce simply to the first term there.  

However, $M_K$ is large not for every $K$ at tree-level where, as can be 
seen in (\ref{MK1to9}), there are two modes $K = 8, 9$ with zero mass.  
Although these modes are expected eventually to acquire masses also from
radiative corrections, they may need special consideration.  Fortunately, 
it turns out that the contributions of these modes 8 and 9 to the amplitude
(\ref{ampegren}) is diagonal in the flavour-states $\tau, \mu, e$ and 
gives therefore no transmutation effects of present interest.  That this
is so can be seen as follows.  We note that in the various terms of the
amplitude (\ref{ampegren}), factors of the following form repeatedly 
appear:
\begin{eqnarray}
\bar{\gamma}_K \bar{\gamma}_K & = \rho^2 |v_K \rangle \langle v_1
   |v_K \rangle \langle v_1|; \ \ \ 
\bar{\gamma}_K^{\dagger} \bar{\gamma}_K^{\dagger} & = \rho^2 |v_1 \rangle 
   \langle v_K|v_1 \rangle \langle v_K|; \nonumber \\
\bar{\gamma}_K \bar{\gamma}_K^{\dagger} & = \rho^2 |v_K \rangle \langle v_1
   |v_1 \rangle \langle v_K|; \ \ \ 
\bar{\gamma}_K^{\dagger} \bar{\gamma}_K & = \rho^2 |v_1 \rangle \langle v_K
   |v_K \rangle \langle v_1|.
\label{gammagamma}
\end{eqnarray}
For $K = 8, 9$, the vectors $v_K$ are orthogonal to $v_1$ so that the two
factors in the first row give zero, while those in the second row are
diagonal in the basis $v_1, v_8, v_9$.  And since $v_1$ is by definition
the vector for the heaviest state $\tau$ while $v_8$ and $v_9$ are 
orthonormal linear combinations of the vectors of $\mu$ and $e$, this 
means that when summed over $8, 9$, the factors in (\ref{gammagamma}) are 
all either zero or diagonal in the flavour states $\tau, \mu, e$.  Hence 
it follows that if one is interested only in non-diagonal processes as we 
are in this paper, then one can omit the terms for $K = 8, 9$ in the sums 
over $K$ above.  

One concludes therefore that to order $s/M_K^2$, off-diagonal (i.e.\  
flavour-violating) amplitudes for the reaction (\ref{egamma}) is given 
just by the first term in (\ref{ampegren}), namely:
\begin{equation}
\bar{u}'(p') \gamma^\mu \frac{i}{(p\llap/ + k\llap/) - m'} \gamma_\mu u'(p)
\label{ampegrena}
\end{equation}
where we recall that $m'$ is the renormalized mass matrix which rotates
with changing scales so that the fermion propagator which depends on $m'$
is no longer diagonal in the lepton-flavour states $\tau, \mu, e$ at the 
energy scale where the reaction is measured.  However, according to 
(\ref{ampegrena}), when evaluating the amplitude, one is to sandwich this 
propagator not between the original lepton-flavour states $u(p)$ but between 
the renormalized states $u'(p)$, which are solutions of the equation 
(\ref{Diracp}) and are thus themselves eigenstates of the renormalized 
(rotated) mass matrix $m'$.  The fermion propagator is thus diagonal between 
these states and give, to order $s/M_K^2$, no non-diagonal matrix elements.  
This is in stark contrast to the effect estimated in \cite{photrans} just 
from the kinemtics of the rotating mass matrix which were quite sizeable.  
In a nutshell, this is because the renormalization mechanism in DSM scheme 
which drives the mass matrix rotation induces at the same time rotations 
in the fermion wave functions and in the interaction vertices, which neatly 
compensate one another to a good approximation to give in the end a near 
null effect.

Similar arguments applied to the reaction (\ref{eebar}) lead to a similar
conclusion.  Of the 1-loop diagrams with (dual colour) Higgs exchange, 
the diagram of Figure \ref{benchfig}(b) is finite and of order $s/M_K^2$.
Then, adding to the tree amplitude (\ref{ampeetr}) the 1-loop diagrams of 
Figure \ref{eediag} replaces the first factor in (\ref{ampeetr}) by:
\begin{eqnarray}
[\bar{u}(p') \gamma^\mu u(p)] & \longrightarrow & 
   [\bar{u}'(p') \gamma^\mu u'(p)] 
   + \frac{\rho^2}{2} [\bar{u}(p') \gamma^\mu (L_L - B_L) u(p)] \nonumber \\
   & & + \frac{\rho^2}{2} [\bar{u}(p') (L_R - B_R) \gamma^\mu u(p)]
\label{ampeeren}
\end{eqnarray}
plus terms involving $\Sigma_c$ and $\Lambda_c^\mu$.  Again for off-diagonal
(i.e.\ flavour-violating) elements, $L_L (L_R)$ cancels with $B_L (B_R)$ 
and $\Sigma_c$ and $\Lambda_c^\mu$ are of order $s/M_K^2$, leaving only the
first term in (\ref{ampeeren}) which has no off-diagonal elements.  The
same arguments hold for the last factor in (\ref{ampeetr}) corresponding 
to the bottom half of the diagrams in Figure \ref{eediag}.  Hence the
conclusion is again that the DSM scheme predicts no flavour-violation in 
the reaction (\ref{eebar}) up to terms of order $s/M_K^2$. 

Recalling our designation at the beginning of the section of transmutation
as flavour-violation due to the rotational effects under changing scales, 
we thus conclude that the DSM scheme predicts no transmutation as such, 
at least at the 1-loop level so far investigated.  And this is the case 
in spite of the rather fast rotation rate the scheme requires for the mass 
matrix in order to explain the observed fermion mixing and mass hierarchy.

There are flavour-violating effects, though not due to rotation, of the 
order of $s/M_K^2$.  But terms of this order would arise in any case, at
least for (\ref{eebar}), from the direct exchange of (dual colour) gauge 
and Higgs bosons.  These (FCNC-type) effects would be common to any model
in which fermion generations are interpreted as a gauged ``horizontal''
symmetry, and not specific to the DSM alone.  The flavour-violation in
such a context is generally taken to be suppressed by large masses for
the exchanged bosons which can be estimated from the experimental bounds 
on flavour-violation.  Specifically, a detailed analysis within the DSM
scheme of meson mass differences and rare meson decays \cite{fcnc} and 
of $\mu-e$ conversion in nuclei \cite{mueconv} led to an estimate of the
gauge boson mass $\mu_N$
of the order $\mu_N/{\tilde g} > 500\ {\rm TeV}$ with $\tilde g$ being
the coupling, which 
is fairly typical for models with ``horizontal'' symmetries.  Although no
similar analysis has yet been performed for the Higgs exchange, a bound
of an analogous order of magnitude is expected for the Higgs mass $M_K$
appearing above, so that at the energy of $\sim 10\ {\rm GeV}$ of present
high sensitivity experiments, the anticipated flavour-violation, in
(\ref{eebar}) for example, will be very small.  Alternatively, one can
turn the considerations around and use the reaction (\ref{eebar}) to set
a bound on the gauge and Higgs boson masses.  Although the bounds so 
deduced are probably going to be less stringent than those obtained from
meson mass differences and rare meson decays even with the data from
new high sensitivity experiments, they will have the virtue of being
free from the many crude assumptions made on the hadron physics inherent 
in the derivation of the bounds with the other methods.

\section{Generalization to Other Cases}

The conclusions on transmutation given in the preceding section have been
deduced explicitly only in the specific DSM scheme as detailed in \cite{ckm}
and applied to two special reactions.  However, the arguments involved such 
as the Ward identity relating wave function and vertex renormalization seem 
quite general and suggest that the result may hold in a more general context.
The results presented above are thus likely to survive some changes in the
details such as the Higgs spectrum of the DSM scheme as given, for example, 
in \cite{ckm, phenodsm}.  This possibility is relevant for although the
DSM as specifically given in \cite{ckm,phenodsm} has so far been remarkably 
successful in reproducing mass and mixing patterns, there may come a point 
when under further detailed examination, minor modifications become necesssary 
for consistency either within the scheme itself or with experimental data 
such as, say, in CP-violation for which the scheme at present says nothing.

The conclusion above of no transmutation for DSM up to terms of orders 
$s/M_K^2$ may also hold for processes other than those two explicitly 
investigated, i.e.\ (\ref{egamma}) and (\ref{eebar}), which are given by 
1-photon exchange.  Although explicit calculations have yet to be performed 
to demonstrate that this is indeed the case, we wish now to explore some 
implications of such a generalization which seem to clarify certain concepts 
we have previously found puzzling.  These conceptual questions would arise 
in any theory with a rotating mass matrix, not just in the DSM scheme alone.

\begin{figure}[ht]
\begin{center}
{\unitlength=1.0 pt \SetScale{1.0} \SetWidth{1.0}
\begin{picture}(150,100)(0,0)
\Line(0,30)(60,30)
\Photon(60,30)(90,60){3}{5.5}
\Line(60,30)(90,0)
\Line(90,60)(100,90)
\Line(90,60)(120,60)
\Text(-10,30)[]{$\mu^-$}
\Text(70,55)[]{$W$}
\Text(110,90)[]{$e^-$}
\Text(130,60)[]{$\bar{\nu}_e$}
\Text(100,0)[]{$\nu_\mu$}
\end{picture} }
\end{center}
\caption{Decay of a $\mu$ giving $\nu_\mu$}
\label{mudecay}
\end{figure}
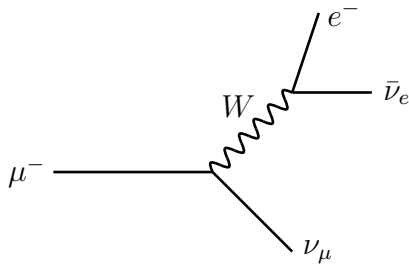

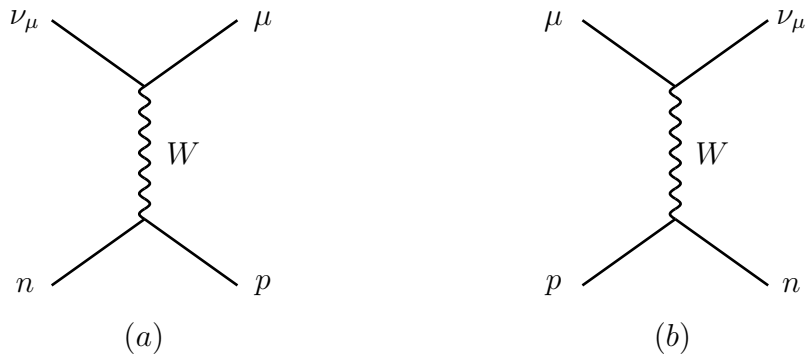
\begin{figure}[ht]
\begin{center}
{\unitlength=1.0 pt \SetScale{1.0} \SetWidth{1.0}

\begin{picture}(350,130)(0,0)

\Photon(65,25)(65,75){2}{6.5}
\Line(65,75)(30,100)
\Line(30,0)(65,25)
\Line(65,25)(100,0)
\Line(100,100)(65,75)

\Text(20,0)[]{$n$}
\Text(20,100)[]{$\nu_\mu$}
\Text(110,100)[]{$\mu$}
\Text(110,0)[]{$p$}
\Text(80,50)[]{$W$}
\Text(65,-20)[]{$(a)$}

\Photon(265,25)(265,75){2}{6.5}
\Line(265,75)(230,100)
\Line(230,0)(265,25)
\Line(265,25)(300,0)
\Line(300,100)(265,75)

\Text(220,0)[]{$p$}
\Text(220,100)[]{$\mu$}
\Text(310,100)[]{$\nu_\mu$}
\Text(310,0)[]{$n$}
\Text(280,50)[]{$W$}
\Text(265,-20)[]{$(b)$}
\end{picture}}
\end{center}
\vspace*{5mm}
\caption{$\nu_\mu$-nucleus collision producing $\mu$ and the time-reversed
process}
\label{munumu}
\end{figure}

A prime example of the sort of questions we wish to pose concerns the 
oscillation of atmospheric neutrinos as reported in \cite{Superk,Soudan}.
Here, what is supposed to have happened is that, say, from $\mu$ decay via 
the process depicted in Figure \ref{mudecay}, one obtains a $\nu_\mu$ which 
is not one of the neutrino mass eigenstates $\nu_i, i = 1,2,3$ but
a linear combination of them, with each having a different mass and 
therefore propagating with a different wave length.  Hence after a while, 
this initial $\nu_\mu$ will no longer remain in a $\nu_\mu$ state but become
a linear combination of $\nu_e, \nu_\mu$ and $\nu_\tau$.  To test whether
the neutrino arriving in the detector is still a $\nu_\mu$ or a linear
combination, what one does is to allow it, for example, to impinge on a 
nucleus, as depicted in Figure \ref{munumu}(a), and see whether a $\mu$ 
is always produced or sometimes an $e$ or $\tau$.  Such a procedure assumes 
of course that the $W$-boson couples always a $\mu$ to a $\nu_\mu$, or 
that, by time-reversal of Figure \ref{munumu}(a), the neutrino produced 
in the $\mu$-nucleus collision of Figure \ref{munumu}(b) is always that 
particular linear combination of the mass eigenstates $\nu_i$ that we 
called $\nu_\mu$.

The last assertion seems obvious until we start to entertain the idea that
mass matrices rotate with changing scales.  The decay process depicted in
Figure \ref{mudecay} occurs at the $\mu$-mass scale $\mu = m_\mu$, while
the production process of Figure \ref{munumu}(b) occurs at a different 
scale depending on the energy.  Given that the mass matrix rotates and
thus  have different orientations and therefore different eigenstates at 
different scales, can we be sure that the neutrino produced in the second 
reaction will always be the same linear combination of $\nu_i$ as that 
obtained from $\mu$-decay?

This question can be unambiguously answered in the present framework by 
repeating the calculation performed above for the reactions (\ref{egamma})
and (\ref{eebar}), namely by evaluating the Feynman digrams with loop
insertions to the tree-diagrams in respectively Figures \ref{mudecay} and
\ref{munumu}(b).  This calculation has not been done.  However, if we assume
that the same result holds here for $W$-exchange as in reactions (\ref{egamma})
and (\ref{eebar}) for $\gamma$-exchange, then the answer to the above question 
is affirmative (up to order $s/M_K^2$), namely that the neutrino obtained 
from the reaction of Figure \ref{munumu}(b) is indeed the same as that 
obtained from $\mu$-decay.  In other words, the neutrino obtained from
$\mu$-decay impinging immediately on a nucleus, i.e.\ without allowing time
for it to oscillate, will produce always a $\mu$ in the final state.  This
is of course the premises on which the experimental analysis is done and
is usually taken as obvious, but in the case where the mass matrix
rotates, it is an assertion which has to be demonstrated, and the above 
argument would now supply the answer.  We note that the same question 
arises not just in the DSM but in principle in any scheme where the mass 
matrix rotates, including in particular the Standard Model as traditionally 
formulated \cite{Ramond,impromat} although there, the rotation being much 
slower than in the DSM, it is not of as much practical significance. 

A similar discussion can be extended to other processes to conclude, for
example, that the $\nu_\mu$ obtained from $\pi$-decay is indeed the same 
as that obtained from $\mu$-decay (again up to order $s/M_K^2$) although
the decays occur in principle at different scales.  Also, by extending
the argument to quarks, similar arguments would lead to the conclusion,
for example, that the CKM mixing element $V_{cb}$ as measured in the 
reaction $b p \longrightarrow c n$ by exchanging a $W$ will be the same
(to order $s/M_K^2$) as the $V_{cb}$ measured in $b$ (i.e.\ D) decay
and independent of scale.

\section{Concluding Remarks}

As in other attempts to explain fermion generations as a ``horizontal
symmetry'' \cite{horizontal}, the DSM scheme necessarily leads to 
flavour-violaton and has to guard against its excessive manifestation.  
A more obvious type of flavour-violation due to exchanges of gauge bosons 
associated with the gauged generation symmetry is easier to guard against 
since this is dependent on the gauge boson mass, usually to its fourth 
power, and so can conveniently be suppressed beyond any given experimental 
bound by assigning to the flavour-changing bosons a sufficiently high mass.  
This genre of flavour-violation in the DSM scheme, as mentioned above, has
already been investigated in some detail \cite{fcnc,mueconv} and it was 
found that by choosing a mass-scale for the associated bosons of
the order of 500 TeV, all existing bounds on flavour-violation can be
satisfied.  There is however in DSM another genre of flavour-violation
which is potentially much more dangerous and which comes about because
the scheme has the, as far as we know, unique feature of explaining
both fermion mixing and fermion mass hierarchy as consequences
of the rotating mass matrix.  The flavour-violation of this genre can
be large depending on the rate at which the mass matrix rotates, and
since this rate is constrained in the scheme by the need to explain the
observed magnitudes of fermion mixing and mass ratios between generations,
there is no adjustable parameter available for tuning the amount of implied 
flavour-violation to escape experimental bounds.  Indeed, it was found
previously \cite{impromat,photrans,transbhar} that judging by kinematics
alone, a mass matrix rotating at the rate the DSM requires can give
flavour-violation of a size readily detectable by modern experiments of 
high sensitivity such as BaBar \cite{Babar} and Belle \cite{Belle} so 
that at one stage we thought we have here a make-or-break test for the 
DSM mechanism.  Hence, the result in this paper that the implied 
flavour-violation is in fact much smaller, though in a sense a 
disappointment, is also a great relief, for otherwise
if experiment finds no 
flavour-violation at the level predicted, there would in principle be no 
escape.  As matters stand, however, the DSM is likely to survive tests 
along these lines for some time to come.

Although little flavour-violation is predicted in the DSM scheme, the 
conventional picture of flavour as a conserved quantity and of different 
flavour states as distinct objects is fundamentally changed.  Flavour 
states rotate into one another so that flavour-violation naturally occurs, 
and though smaller than naively expected, flavour-violation is nevertheless 
present and in principle detectable.  Indeed, the effects expected are 
similar in magnitude to flavour-changing neutral current effects 
\cite{fcnc,mueconv} and can possibly be observable soon by experiment under 
certain circumstances.  Besides, in vector boson decays as studied in 
\cite{transbhar} where the $B_L, B_R$ and $L_L, L_R$ terms from respectively 
the wave function and vertex renormalization may not cancel exactly
for lack 
of a Ward identity, flavour-violation need not be also of order $s/M_K^2$ 
and hence may be detectable already by current experiments \cite{Bepc,Cleo}.

For the present moment, however, the result of this paper seems to allow 
schemes like the DSM to both ``have the cake and eat it'', i.e.\ both to 
explain the sizeable fermion mixings and mass ratios between generations 
by a mass matrix rotating at appreciable speed, and at the same time
to avoid contradiction with experiment as regards flavour-violation.  
Besides, as explained in the preceding section, it helps to resolve some 
conceptual difficulties concerning the definition of mixing matrices and 
neutrino oscillations when the mass matrix rotates.

Within the DSM framework, the present paper represents also 
a certain technical step forwards in that previous studies of
scale-dependent renormalization effects have been limited to
only single-particle properties
such as masses and mixing angles, but have now been extended to two-body 
properties observable only in collisions.

\end{document}